# Intra-Cluster Autonomous Coverage Optimization For Dense LTE-A Networks

Ali A. Esswie
Network Performance Group, North Africa RSRC, Huawei Technologies.

*Abstract*— Self Organizing Networks (SONs) are considered as vital deployments towards upcoming dense cellular networks. From a mobile carrier point of view, continuous coverage optimization is critical for better user perceptions. The majority of SON contributions introduce novel algorithms that optimize specific performance metrics. However, they require extensive processing delays and advanced knowledge of network statistics that may not be available. In this work, a progressive Autonomous Coverage Optimization (ACO) method combined with adaptive cell dimensioning is proposed. The proposed method emphasizes the fact that the effective cell coverage is a variant on actual user distributions. ACO algorithm builds a generic *Space-Time* virtual coverage map per cell to detect coverage holes in addition to limited or extended coverage conditions. Progressive levels of optimization are followed to timely resolve coverage issues with maintaining optimization stability. Proposed ACO is verified under both simulations and practical deployment in a pilot cluster for a worldwide mobile carrier. Key Performance Indicators show that proposed ACO method significantly enhances system coverage and performance.

*Keywords*— SON; Coverage; Optimization; LTE-A; AoA; FDD; Autonomous

## I. INTRODUCTION

As per the emerging need for complex Long Term Evolution - Advanced (LTE-A) networks, it is becoming one the most critical challenges against worldwide mobile operators to provide continuous network coverage with better mobile services, while reducing the management and maintenance costs [1]. And towards the upcoming densely networks, human involved optimization not only increases the Operational Expenditures (OPEX) extensively, but also becomes hardly feasible due to the unprecedented network complexity with increasing number of radio parameters. Hence, Self Organizing Network (SON) had been part of the 3GPP standardization since release 9 and it aims to enabling network autonomous optimization [2, 3].

Through the coverage optimization process, there are several issues that need to be timely resolved to maintain the standard Quality of Service (QoS) such as blind spots due to antenna scan losses, extended or limited downlink coverage, and mismatch between Uplink (UL) and Downlink (DL) coverage. Though, the DL coverage optimization is the major source of interest in this paper because of its major impact on whole network QoS. The DL coverage optimization process usually consists of two phases, initial coverage planning and coverage maintenance. The former is performed using vendor specific planning tools which are unable to adapt to complex networks due to their modeling limitations. The latter requires many Drive Tests (DTs) that consequently add much higher input costs.

In literature, Adjusting the transmission power of the eNodeB is always the most common way for coverage optimization [4, 5]. However, employing only power adjustment for coverage optimization is sub optimal since when the power adjustment limits are reached, persisting abnormal coverage can't be optimized. Moreover, heuristic schemes are proposed based on Golden Section Search (GSS) [6]. In [7], authors proposed a centralized system that adaptively sets the global network parameters to ensure overall coverage. And to mitigate signaling overhead, authors in [8] extended the proposal in [7] to a distributed nature to be run on cell level. The majority of the research contributions didn't put a major attention to the practicality of the proposed schemes in the real complex environments. The majority of the research contributions didn't put a major attention to the practicality of the proposed schemes in the real complex environments, disregarding the critical requirements of the mobile operators in terms of special geographical network capacity, VIP traffic handling, traffic maximization and offloading between different air interfaces e.g. 3G U1, U2, U9 and 4G.

In this paper, a novel progressive Autonomous Coverage Optimization (ACO) algorithm is proposed. ACO is executed per site cluster in order to provide a generic framework towards mobile operator requirements while avoiding ping-pong optimization. It projects the received user measurements into a Space – Time virtual map that is adaptively generated for each cell in each cluster based on planning data. ACO allows for a generic diversity of coverage optimization precisions across network clusters according to traffic & user priorities. A progressive manner of optimization with memory states is adopted by ACO to timely resolve the faulty coverage spots without incurring larger delays by enforcing incremental aggressive optimization actions with time. ACO is verified under MATLAB simulations and a practical deployment in a commercial 4G cluster as well. In both cases, ACO shows significant improvement in network performance and coverage.

This paper is organized as follows. Section II presents the system model in this work. Section III introduces full details of the proposed ACO. Section IV shows the performance evaluation metrics of ACO in both cases: Matlab simulations and the practical deployment. Finally, the conclusion is drawn in the last section.

## II. SYSTEM MODEL

An FDD LTE downlink with $C = \{c_1, c_2, \ldots, c_C\}$ cells, each antenna sector is $120^o$ and $U = \{u_1, u_2, \ldots, u_U\}$ users per cell is considered in this work as in Fig. 1. Another network software element entitled as *eOptimizer (eOpt)* is deployed over which the ACO algorithm runs. For the sake of practical compatibility, *eOpt* is deployed on the same vendor-specific server which originally is connected to all Network Elements (NEs) for reporting the periodic KPIs.

The cell specific Reference Signals (RS) are all QPSK modulated – a constant modulus modulation. The power of the RS is $P_{RS}$ and bounded by $[0, P_{RS,max}]$. The RS waveform can be written as:

$$R_{l,n_s}(m) = \frac{1}{\sqrt{2}}[1 - 2c(2m)] + j\frac{1}{\sqrt{2}}[1 - 2c(2m+1)] \quad (1)$$

where $m$ is the RS index, $n_s$ is slot number within the current radio frame and $l$ is the symbol index. $c(i)$ is pseudo-random sequence of length-31 gold sequence.

Assuming proper planning of the network Primary Cell Id (PCI) in terms of the *modulo order,* where unit direction cells should be assigned unique $PCI\ modulo\ n$ and $n$ is the reuse factor, the RS signals can be only affected by minimum interference due to the PCI reuse and RS Signal to Interference Noise Ratio (SINR) can be written as in [9]:

$$SINR_{RS} = \frac{P_{RS} H_{mc} G_{mc}}{\sigma^2 + \sum_{d=I_c} P_d H_{md} G_{md}} \quad (2)$$

where $H_{mc}$ & $G_{mc}$ are the channel and antenna gains between cell $c$ and user $m$ respectively and $I_c$ is all interfering cells on the current cell RS waveform.

### III. PROPOSED AUTONOMOUS COVERAGE OPTIMIZATION (ACO)

The proposed ACO algorithm is run per every cluster on *eOpt*, which is a centralized software collocated on the same vendor specific server that is originally connected to all NEs.

Hence, ACO doesn't add additional input costs or communication overhead. *eOpt* is initialized by two periodicities: $t_{MR}\ and\ t_{eOPT}$. The former is the time periodicity to import fed-back user Measurement Reports (MRs) into *eOpt* and it is a fixed period to increase or decrease data precision. While the latter is the time periodicity of the algorithm to run and it is adaptively set based on traffic volume and user number to ensure proper optimization. *eOpt* is also set with the preplanned Engineering Parameter Table (EPT) of all contributing cells for MRs mapping. Each cell in the network contributes by localizing its serving users' MRs in order for the *eOpt* to project them on its virtual coverage map.

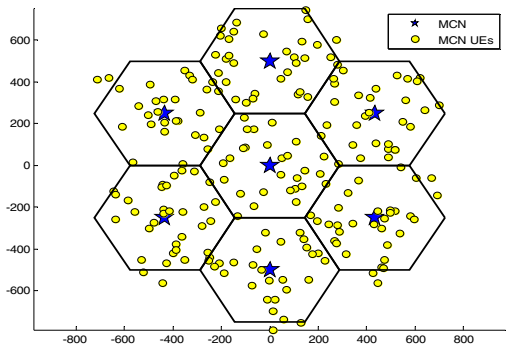

Fig. 1. System model: 3 cells/ site, each is $120^o$ of coverage

MR localization information's are extracted from the users' UL transmissions through the Angle of Arrivals (AoAs) and difference in Time Advance (TA). Consequently, it requires no additional signaling overhead. Then, *eOpt* virtualizes all MRs from all cells in every cluster on a 2D coverage map / cell. The structure of the coverage mapping is arbitrary. For example as in Fig. 2, the coverage distance is divided non uniformly into 3 sub-areas with a minimum distance of $d$ meters. Non uniformity is useful to change coverage optimization precision at any specific region of the cell or cluster. Angle sub-areas are proposed to be progressive with distance in order to increase the angle optimization precision where wide angles exist. Hence, the MRs' localization error is minimized as long as it is limited within one sub-area on the virtual coverage map.

Then, *eOpt* projects MRs received per cell on the shown Space - Time map, where all MRs are projected to the sub-area where they were originally reported from (Angle – Distance) and then, all data assigned to each sub-area is averaged over all times bounded by the previous $t_{MR}$ period. By each $t_{eOPT}$, The ACO algorithm runs. On each cell, traffic volume and number of users should be proper for the ACO to proceed running without experiencing ping-pong optimizations. ACO firstly detects overshooting or limited coverage conditions from user measured TAs. Afterwards, ACO proceeds to adjust the actual cell coverage to match the forecasted user distribution over the passed $t_{MR}$. Progressive levels of optimization are followed based on a sliding window to faster algorithm convergence as it will be discussed later.

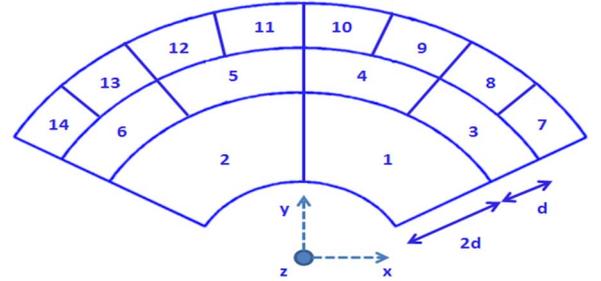

Fig. 2. Coverage Mapping/cell over angle-distance sub-areas

### A. ACO Design Parameters

In this section, the major design parameters of the ACO algorithm are presented as follows.

- $t_{MR}$ : the periodicity of importing user MRs into *eOpt*.

- $t_{eOPT}$ : the periodicity of the ACO algorithm to run, which is set adaptively based on the proper traffic volume.

- $W$: the optimization sliding window size, where incremental actions are considered based on the action history within the current window.

- $n_d$ & $n_\theta$ : the arbitrary number of distance and angle sub-areas on the coverage map respectively.

- $\tau_{RSRP}$, $\tau_{Traffic}$ & $\tau_{User}$ : thresholds for average RSRP, traffic volume and number of RRC connected users per cell that should be satisfied in order for ACO to run properly.

### B. MRs Projection on The eOpt Coverage Map

A predefined coverage is set independently for each cell in each cluster based on the initial planning data and it can be expressed as shown in Fig. 3,

$$\theta_{tilt} = \theta_{geo} + \theta_{VerB} \quad (3)$$

$$R_{exp} = \frac{(H_{BS} - h_{MS})}{\tan(\theta_{tilt})} \quad (4)$$

$$d = \frac{R_{exp}}{n_d} \quad (5)$$

where $\theta_{geo}$ is the mechanical tilt angle, $\theta_{VerB}$ is the vertical beamwidth, which corresponds to the electrical antenna tilt. $H_{BS}$ and $h_{MS}$ are the heights of the BS and UE respectively. $R_{exp}$ is the predefined ideal coverage of each cell and hence $d$ is the minimum distance space for a sub-area.

And due to the fact that equation (4) overestimates the downtilt angle in such way that extensively increases the MRs localization error, a more empirical relation is derived:

$$G_{BS}(\theta) = 3\left(\ln(H_{BS} - R_{exp}^{0.8})\right) \times \log_{10} \theta_{VerB} \quad (6)$$

From the *eOpt* perspective, the final coverage map per cell is a group of sub-areas in terms of start & end angle in addition to a total ideal coverage distance. *eOpt* accordingly extracts location information from user MRs and assigns them to corresponding sub-areas.

Angle information is extracted from user UL sounding through the Angle of Arrivals (AoAs). UL AoAs are estimated based on the efficient and hardware friendly Capons' estimation scheme where the beam is originally formed across the angular region of interest and the angle that provides the highest power is the estimated 1st AoA.

The Capon spatial power spectrum is characterized by:

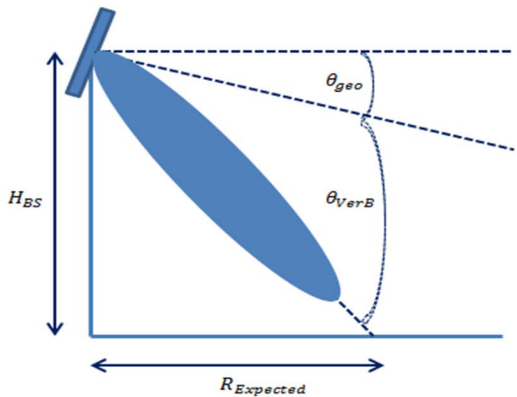

Fig. 3. Expected Ideal Coverage / Cell

$$P_{Capon}(\emptyset) = \frac{1}{a^H(\emptyset) R_{xx}^{-1} a(\emptyset)}$$

$$\emptyset^o = \arg\max(P_{Capon}(\emptyset)) \quad (7)$$

$$R_{xx} = \frac{1}{N} \sum_{1}^{n} x(n)x^H(n)$$

where $R_{xx}$ is the $M \times M$ auto covariance matrix and $a(\emptyset)$ is the $M \times 1$ steering vector that finally points to the azimuth angle $\emptyset^o$, where $M$ is the number of transmit antennas. The 3GPP spatial channel from the $s_{th}$ to $u_{th}$ antenna on the $n_{th}$ channel sub path is given by:

$$h_{u,s,n}(t) = \sqrt{\frac{P_n \sigma_{sf}}{M}} \sum_{m=1}^{M} \left( \sqrt{G_{BS}(\theta_{n,m,AoD})} \; e^{j[kd_s \sin(\theta_{n,m,AoD}) + \overbrace{\varphi_{n,m}}^{\text{Random Phase}}]} \right.$$
$$\left. \cdot \sqrt{G_{MS}(\theta_{n,m,AoA})} \; e^{j[kd_u \sin(\theta_{n,m,AoA})]} \; e^{jk\|v\| \underbrace{\cos(\theta_{n,m,AoA} - \theta_v)}_{\text{Doppler}} t} \right) \quad (8)$$

Hence, the UL and DL antenna response coefficients are dependent on the frequency gap between UL and DL links:

$$\mathbf{a}_{UL}(\vartheta) = \begin{bmatrix} 1 \\ e^{-j2\pi\Delta \frac{f_{UL}}{F_0} \sin(\vartheta)} \\ \vdots \\ e^{-j2\pi\Delta \frac{f_{UL}}{F_0}(N-1)\sin(\vartheta)} \end{bmatrix} \quad \mathbf{a}_{DL}(\vartheta) = \begin{bmatrix} 1 \\ e^{-j2\pi\Delta \frac{f_{DL}}{F_0} \sin(\vartheta)} \\ \vdots \\ e^{-j2\pi\Delta \frac{f_{DL}}{F_0}(N-1)\sin(\vartheta)} \end{bmatrix}$$

$$\mathbf{a}_{DL}(\vartheta) = \mathbf{T}(\vartheta)\mathbf{a}_{UL}(\vartheta), \quad (9)$$

$$\mathbf{T}(\vartheta) = \text{diag}(1, e^{j2\pi\Delta \frac{f_{UL} - f_{DL}}{F_0} \sin(\vartheta)}, \ldots, e^{j2\pi\Delta \frac{f_{UL} - f_{DL}}{F_0}(N-1)\sin(\vartheta)})$$

Consequently, for FDD systems, UL AoAs adjustment scheme is required due to the frequency shift between UL and DL. The physical fact that UL and DL waves bounce along the same set of clusters to their destinations, given a relatively low frequency gap, has led to implementing a simple linear transformation to transform UL signatures into DL ones although the exact transformation is non linear [10]. As clear in Fig. 4, increasing the frequency gap between UL and DL channels leads to more deviation among UL and DL directions and hence to a poor DL localization. Once the UL principal directions are estimated from (7), the final DL steering vector can be generated to the antenna elements.

$$AoD_{dl} = \Phi \, AoA_{ul} \quad (10)$$

$$\Phi = A_{dl} A_{ul}^H (A_{ul} A_{ul}^H)^{-1} \quad (11)$$

where $\Phi$ is the linear transformation matrix estimated from the quantized fed-back knowledge of the downlink channel $A_{dl}$ and estimated uplink channel from UL sounding signals $A_{ul}$.

## C. Autonomous Performance Evaluation and Optimization

By every $t_{MR}$ period, the ACO algorithms starts evaluating each cell coverage conditions whether it requires further optimization or not as follows.

$$RSRP_{>75\%} \leq \tau_{RSRP}$$

where the $\tau_{RSRP}$ is the set threshold that 75% (for coverage probability of 0.75) of the cell coverage CDF should satisfy. Practically, it is set to -80dBm for standard coverage.

The traffic volume and the number of users should initially satisfy predefined thresholds to avoid ACO ping-pong optimization:

$$v_{traffic} \geq \tau_{Traffic}$$
$$n_{user} \geq \tau_{User}$$

If previous conditions are satisfied, the ACO evaluates the cell coverage conditions by investigating the measured TAs for each cell as given by:

$$R_{md,avg} > \partial R_{exp}, \quad \partial > 1$$

where $R_{exp}$ & $R_{md,avg}$ are the predefined and measured coverage of the cell and $\partial$ is scaling factor. The latter is estimated based on measured Time Advance (TA) as follows.

$$TA_{DL} = eNB_{RX,TX} - UE_{RX,TX} \quad (12)$$

where $eNB_{RX,TX} = eNB_{RX} - eNB_{TX}$ is the time difference between uplink reception and downlink transmission of subframe #i. The minimum granularity of the TA is $16T_s$, where $T_s$ is sampling time. Hence, one TA sample corresponds to a coverage of 78 meters. If condition is satisfied, cell is identified as overshooting. Hence, a down-tilt is needed:

$$\theta_{DT} = floor\left(tan^{-1}\left[\frac{(H_{bs} - h_{ms})}{2d'}\right]\right)$$
$$d' = R_{md,avg} - R_{exp} \quad (13)$$
$$\theta_{DT,new} = \theta_{DT,current} + \theta_{DT}$$

where $\theta_{DT,new}$ is the updated cell down-tilt angle for. If (13) is not possible where tilt angle is at maximum, ACO steps down to next action through the careful reduction of the RS power:

$$P_{RS,new} = P_{RS,current} - \Delta \quad (14)$$

Otherwise if:

$$R_{md,avg} < \varepsilon R_{exp}, \quad \partial \gg \varepsilon, \quad \varepsilon < 1$$

Then, the cell is identified as a limited coverage cell. ACO searches for the worst coverage sub- areas and generates a Top Worst Area (TWA) list. First, it detects the location of the majority of users who experienced faulty coverage as follows.

$$R_{md,\gamma\%} < \frac{R_{exp}}{2} ?$$

where $R_{md,\gamma\%}$ is the $\gamma^{th}$ percentile of the measured coverage CDF from user reported TAs.

**OK**: a down-tilt is applied and $\theta_{DT}$ is updated as in (6) with:
$$d' = \frac{R_{exp}}{4}$$

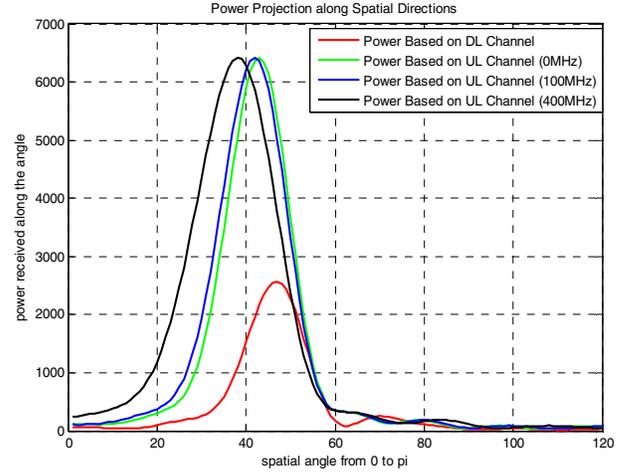

Fig. 4. UL & DL AoAs Correspondence

**NOK**: an up-tilt is applied and $\theta_{UT}$ is updated as in (6) with:

$$d' = \frac{R_{exp}}{2} + \frac{R_{exp}}{4}$$

This way, ACO biases the coverage optimization to the faulty coverage locations where the majority of users exists. Otherwise, in case of diverse user distribution, ACO searches for where the major faulty coverage exists.

$$C_a = count\left(j(<x) > \frac{R_{exp}}{2}\right), C_b = count\left(j(<x) < \frac{R_{exp}}{2}\right)$$

where $j(<x)$ is the sub-area index in where the average RSRP is less than the predefined threshold $x$.

$$if\ C_a > C_b$$

**OK**: an up-tilt is applied and $\theta_{UT}$ is updated as in (6) with:

$$d' = \frac{R_{exp}}{2} + \frac{R_{exp}}{4}$$

**NOK**: a down-tilt is applied and $\theta_{DT}$ is updated as in (6) with:
$$d' = \frac{R_{exp}}{4}$$

In all cases, the final down-tilt angle will be updated as:
$$\theta_{DT,new} = \theta_{DT,current} \pm \frac{\theta_{DT}}{\theta_{UT}} \quad (15)$$

And when down-tilt or up-tilt actions are not possible, the next level of optimization is applied:

$$P_{RS,new} = P_{RS,current} \pm \Delta \quad (16)$$

However, the majority of users may be distributed over the angle dimension, rather than only the distance dimension. ACO rotates the antenna excitement coefficients to match the location of the estimated DL AoDs extracted from UL AoAs as in (10) and the final steering vector can be given as:

$$a(\emptyset) = \left[1, e^{\rho d cos\emptyset'}, e^{\rho 2d cos\emptyset'}, \ldots\ldots, e^{\rho(M-1)d cos\emptyset'}\right]^T$$

$$\rho = \frac{2\pi}{\lambda} \quad \& \quad \emptyset' = \emptyset_{DL} \pm \epsilon$$

Where the $\epsilon$ is the beam rotation angle. The angular space per cell is divided into two major sub-areas, each is of an angle coverage of $60^o$ [$0\ to\ 60^o\ and\ 60\ to\ 120^o$], indicated by the central vertical line in Fig. 2. If an aggressive majority of the estimated user AoDs are located in a single sub-area, a rotation angle is applied and data beams are shifted to either right or left to enhance the angular coverage precision where most of the users exist. The final shot beams are shown in Fig. 5.

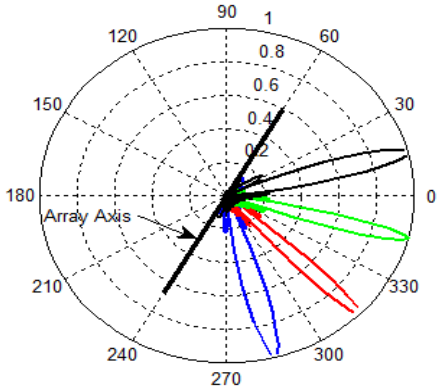

Fig.5. Base Beams with equal AoDs directions for sector $i, \epsilon = 0$

*D. Progressive Optimization*

ACO adopts a progressive level of optimization actions through a sliding window of size $W$ over time. This way, ACO reaches out convergence faster by adopting incremental actions with time given that former actions hadn't satisfied the predefined conditions:

For example, as in Fig.6, for a sliding window of size $w = 3$, when two successive down-tilts or up-tilts are occurred over $w-1\ \&\ w-2$, a progressive reduction of the RS power is adopted in the current $w$ running period to foster the algorithm convergence. By this, ping pong optimization is avoided along with lower convergence delay.

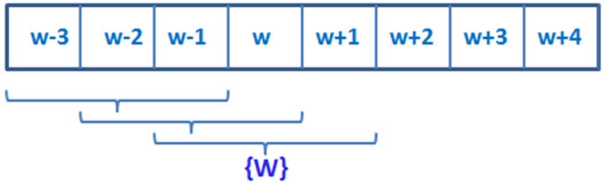

Fig. 6. Sliding progressive window of size W=3

IV. PERFORMANCE EVALUATION

In this section, the performance evaluation of the proposed ACO algorithm is presented. ACO is verified under MATLAB simulations in addition to practical deployment on a worldwide mobile carrier commercial cluster, which is initially well preplanned. In both cases, ACO showed significant enhancement in cluster coverage and accordingly the experienced throughput. Table I shows the major simulation parameters.

In Fig. 7, ACO shows significant improvement of the cluster coverage over only 2 rounds of autonomous optimization (4 days). As clearly shown, the major coverage enhancements are towards the areas associated with very bad coverage levels or no coverage where they are timely detected as coverage holes.

TABLE I: SIMULATION PARAMETERS

| Parameter | Value |
| --- | --- |
| Channel model | 3GPP 3D model |
| Deployment | Homogenous Macro |
| Network | Wraparound, 21 cells |
| UE dropping | 2D UE Dropping, 10 UEs/Cell |
| eNodeB antenna configurations | $4 \times 1, 8 \times 1$ |
| eNodeB antenna polarization | ULA, 0.5 $\lambda$ |
| UE antenna configuration | $2 \times 1$ |
| UE antenna polarization | ULA, 0.5 $\lambda$ |
| TX Mode | MIMO (TM3) |
| $d$ | 210 meters |
| $t_{MR}$ | 15 minutes |
| $t_{eOPT}$ | Initially 2 days |
| $w$ | 3 |
| $n_d\ \&\ n_\theta$ | 4 & 2 – 4 – 8 |
| $\tau_{RSRP}, \tau_{Traffic}\ \&\ \tau_{User}$ | -80dBm, 25GB & 9 UEs/Cell |
| $\partial, \varepsilon, \Delta$ | 2.1, .4 & 1dBm |
| $\epsilon$ | $15^o$ |

The cell aggregate throughput is improved, especially in the areas of the recently optimized coverage levels due to the reduction of the Block Error Rates (BLER), resulted from the RRC_CONN_RECFG abnormal releases, when the signal and noise levels are approximately same as in Figs 8 & 9.

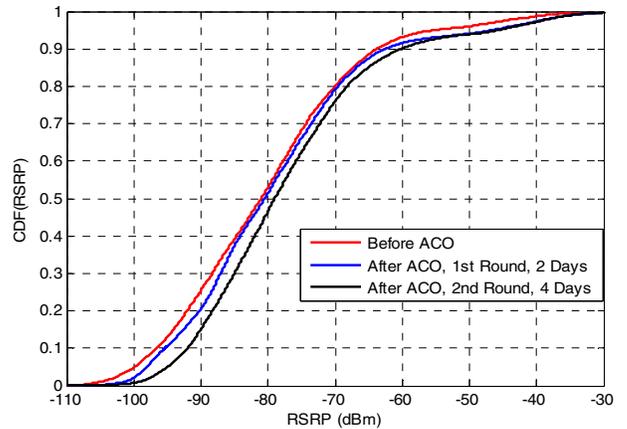

Fig. 7. RSRP Comparison over ACO: 2 Iterations

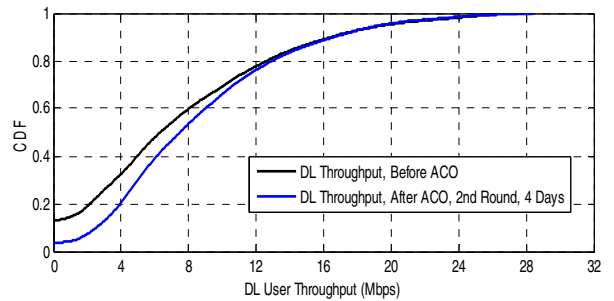

Fig. 8. Throughput Comparison over ACO: 2 Iterations

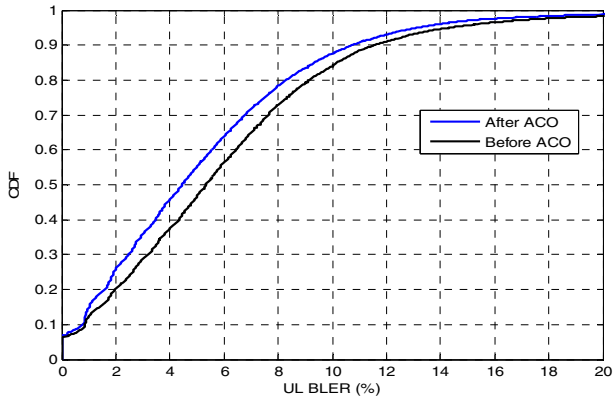

Fig. 9. BLER Comparison over ACO: 2 Iterations

Additionally, ACO is deployed practically on a well-preplanned commercial 4G cluster of 32 sites in a major city with mature traffic and user capacity. As clearly shown in Figs. 10 & 11, ACO detected the coverage gaps with passing iterations by applying corrective actions. This way, ping pong optimization is effectively avoided when specific actions are applied in one iteration and being rolled back the successive one in a way that results in an increased optimization delay. ACO showed an extensive practicality in real-time networks.

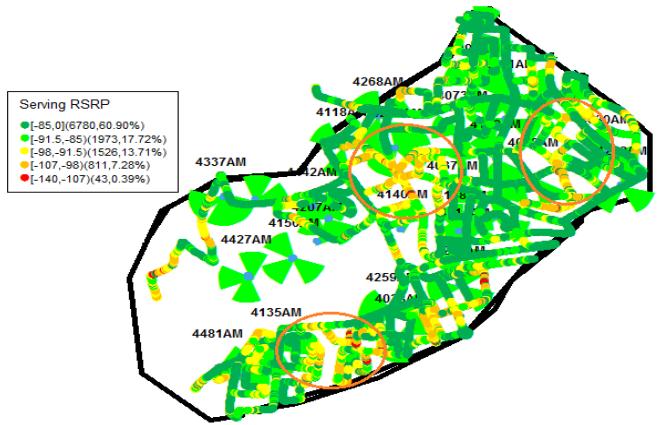

Fig. 10. Practical ACO Deployment: Drive Test Before ACO

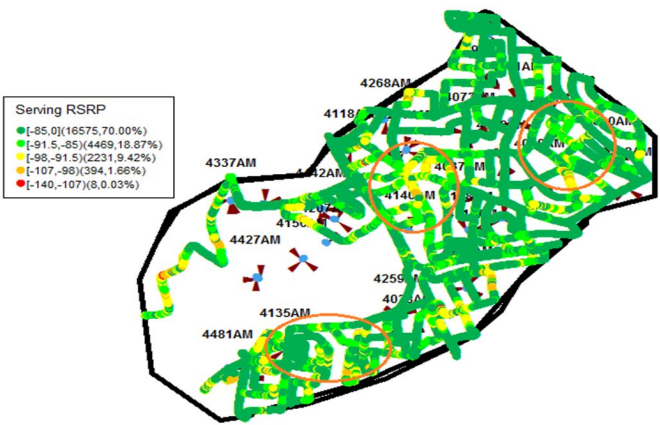

Fig. 11. Practical ACO Deployment: Drive Test After ACO

## V. CONCLUSION

In this paper, an Autonomous Coverage Optimization (ACO) method was proposed. ACO algorithm required no additional communication overhead or input costs. It projected conventional user measurement reports on a predefined space-time coverage map per cell in terms of an angle and a distance. Corrective RF actions are applied in A progressive manner and final steering vectors are adaptively adjusted to maintain standard coverage levels in the areas where the majority of users exists. ACO is verified under MATLAB simulations in addition to practical deployment in a commercial cluster for worldwide mobile carrier. In both cases, ACO showed clear coverage and quality improvements with maintaining optimization stability and avoiding optimization ping-pongs. for the latter case, ACO had shown an extensive compatibility to real time networks with coverage gain of ~10% in the -85dBm coverage region.